\documentclass[conference]{IEEEtran}

\ifCLASSINFOpdf
\else
\fi
\usepackage{graphicx}
\usepackage{subfigure}
\usepackage{psfrag}
\usepackage{amsmath}
\usepackage{amsfonts}
\usepackage{amssymb}
\usepackage{float}

\newtheorem{definition}{Definition}

\newtheorem{lemma}{Lemma}

\makeatletter
\def\ifundefined{\@ifundefined}
\makeatother

\begin{document}
%
\title{Outage Capacity of Incremental Relaying \\ at Low Signal-to-Noise Ratios}
\author{\IEEEauthorblockN{T. Renk\IEEEauthorrefmark{1}, H. J{\"a}kel\IEEEauthorrefmark{1}, F.~K. Jondral\IEEEauthorrefmark{1}, D. G\"{u}nd\"{u}z\IEEEauthorrefmark{2}\IEEEauthorrefmark{3}, A. Goldsmith\IEEEauthorrefmark{3}}\\
\IEEEauthorblockA{
            \IEEEauthorrefmark{1}Institut f{\"u}r Nachrichtentechnik, Universit{\"a}t Karlsruhe (TH) \\}
\IEEEauthorblockA{
            \IEEEauthorrefmark{2}Dept.\ of Electrical Engineering, Princeton University, Princeton, NJ \\}
\IEEEauthorblockA{
            \IEEEauthorrefmark{3}Dept.\ of Electrical Engineering, Stanford University, Stanford, CA \\}

  Email: \{renk,jaekel,fj\}@int.uni-karlsruhe.de, dgunduz@princeton.edu, andrea@wsl.stanford.edu,
  \thanks{}
}


\def\snr{{\sf SNR}}
\def\dB{{\sf dB}}
\def\outcap{$\epsilon$-outage capacity }

\maketitle

\begin{abstract}
We present the \outcap of incremental relaying at low signal-to-noise ratios (SNR) in a wireless cooperative network with slow Rayleigh fading channels. The relay performs decode-and-forward and repetition coding is employed in the network, which is optimal in the low SNR regime. We derive an expression on the optimal relay location that maximizes the $\epsilon$-outage capacity. It is shown that this location is independent of the outage probability and SNR but only depends on the channel conditions represented by a path-loss factor. We compare our results to the \outcap of the cut-set bound and demonstrate that the ratio between the \outcap of incremental relaying and the cut-set bound lies within $1/\sqrt{2}$ and $1$. Furthermore, we derive lower bounds on the \outcap for the case of $K$ relays.\newline

\end{abstract}


%
\IEEEpeerreviewmaketitle

\section{Introduction}\label{sec:intro}
\PARstart{C}{ooperation} in wireless networks is a promising technique to mitigate fading, which results in a fluctuation in the amplitude of the received signal. The basic idea behind cooperation is that several users in a network pool their resources in order to form a `virtual' antenna array which creates spatial diversity \cite{sendonaris03,sendonaris03_2,renk07_3}. This diversity leads to an increased exponential decay rate in the error probability with increasing signal-to-noise ratio (SNR) and thus becomes more evident in the high SNR regime. For instance, a diversity order of $2$ means that the outage probability decreases proportional to $10^{-2}$ with a $10$ dB increase in the SNR of the system \cite{laneman04}. However, SNR cannot be increased arbitrarily. Especially for applications such as ad-hoc and sensor networks, power (or energy) plays a major role in the design since it is a limited resource \cite{goldsmith02}. So, for practical considerations, the low SNR regime is much more interesting.

\begin{figure}[t]
\centering
\psfrag{A}{$\rm S$}
\psfrag{B}{$\rm R$}
\psfrag{C}{$\rm D$}
\psfrag{a}{$h_{\rm sr}$}
\psfrag{b}{$h_{\rm sd}$}
\psfrag{c}{$h_{\rm rd}$}
\psfrag{d}{BC}
\psfrag{e}{MAC}
\includegraphics[width=0.45\textwidth]{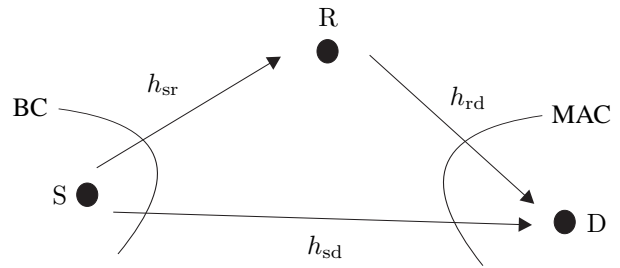}
\caption{Network consisting of source $\rm S$, relay $\rm R$, and destination $\rm D$. Channel gains are represented by $h_{\rm sd},h_{\rm sr}$, and $h_{\rm rd}$. BC and MAC denote the cuts for the broadcast channel and the multiple access channel, respectively.}
\label{fig:network}
\end{figure}

For delay constrained transmission over slowly varying channels, the metric of pure capacity (`Shannon' capacity) which is the maximal transmission rate for which the error probability can be made arbitrarily small (under an average power constraint) is not useful anymore. In the strict sense, the Shannon capacity for those channels is $0$ and other metrics have to be found that are more suitable. Therefore, \outcap has been defined as the maximal transmission rate for which the outage probability is not larger than $\epsilon$ \cite{ozarow1994, tse2005}. The reason why outage probability is considered here is due to the fact that it approximates the error probability quite well in coded systems with long enough block size \cite{caire1999}. In \cite{avestimehr07} \outcap in the low SNR regime has been investigated for a frequency division cooperative system. It is shown that a bursty amplify-and-forward protocol achieves the optimal performance and that the \outcap for the non-coherent scenario is equal to the coherent one. A scheme that can also have a variable transmission rate dependent on the channel conditions has been investigated in \cite{jindal08}. The authors considered the performance of hybrid automatic repeat request (ARQ) where the data rate is increased in order to maintain a constant outage probability.

In this paper, we consider incremental relaying (IR) \cite{laneman04}. In IR networks, the destination sends a one-bit acknowledgment (ACK) to the relay and the source if it is able to decode the source message successfully. If not, it sends a NACK to indicate failure of transmission. In this case the relay, if it has been able to decode the source message, forwards the source message to the destination by employing repetition coding. The destination then performs maximum ratio combining of the signals from the source and the relay, which leads to an accumulation of SNR. We stress that by using jointly designed but independent codebooks (i.e., parallel channel coding), it is generally possible to achieve better results. However, in the low SNR regime, parallel channel coding can be deduced to repetition coding which shows that repetition coding is optimal \cite{avestimehr07}.

The remainder of the paper is organized as follows. In Section \ref{sec:net_mod} we present the system model. Section \ref{sec:out_cap} deals with the $\epsilon$-outage capacity. We first introduce the incremental relaying protocol and derive the corresponding expression on $\epsilon$-outage capacity. Then we consider the cut-set bound in the case of a single relay. The optimal relay location where \outcap is maximized is also presented. In Section \ref{sec:comp} we compare the \outcap of incremental relaying to the cut-set upper bound and generalize our results to the case of $K$ relays. Finally, Section \ref{sec:con} concludes the paper.


\section{Network Model}\label{sec:net_mod}
We consider the network depicted in Fig.~\ref{fig:network} which consists of a source $(\rm S)$, a relay $(\rm R)$, and a destination $(\rm D)$. The channel gains $h_{i}$, $i \in\{{\rm sd},{\rm sr},{\rm rd}\}$, are from a slow Rayleigh fading profile with variances $\sigma_i^2$. Hence, $|h_{i}|^2$ follows an exponential distribution with mean value $\sigma_{i}^2$ and phases are uniformly distributed over $[0,2\pi)$. A common path-loss model is applied, where the channel variances $\sigma_{i}^2$ are proportional to $d_{i}^{-\alpha}$ with $d_{i}$ being the distance between two nodes. The parameter $\alpha$ describes the path-loss exponent which typically lies between $3$ and $5$ for cellular mobile networks. We also have white Gaussian noise added at each receiving node. Noise realizations are considered to be independent and identically distributed (i.i.d.) and all come from a zero-mean Gaussian distribution with variance $N$, i.e., $n \sim {\cal CN}(0,N)$. An average transmit power constraint of $P$ is assumed at the source and the relay over a transmission block and SNR is defined by $\snr=P/N$. A practical constraint is imposed on the relay which allows the relay to either receive or transmit at any instant, but not to do both simultaneously, i.e., the relay operates in the half-duplex mode. Moreover, we assume that the relay employs decode-and-forward protocol for cooperation, which means that the relay decodes the source message and encodes it again before retransmission. This relay strategy has the advantage that there is no noise enhancement compared to amplify-and-forward, where the relay simply amplifies its receive signal with a certain amplification factor to satisfy the power constraint. We assume throughout the paper that there are enough channel uses per transmission phase so that the codes achieve their intended rates reliably if they are below the channel capacity.


\section{Outage Capacity}\label{sec:out_cap}

\subsection{Definition}
As already mentioned before, the Shannon capacity in a slow fading channel is $0$ when the transmitter does not have channel state information. Therefore, \outcap has been introduced as a new performance metric \cite{ozarow1994}. It is defined as follows:

\vspace*{0.2cm}
\begin{definition}\label{def:out_cap}
$\epsilon$-outage capacity $C_{\epsilon}$ is the highest rate $R$ such that outage probability satisfies $p_{\rm out}(R, \snr) := \Pr(C(\snr) < R) \leq \epsilon$, where $0 \leq \epsilon \leq 1$ and $C(\snr)$ is the instantaneous capacity, which is a random variable due to random variations in the channel. For a given $\epsilon$, we have:
\begin{equation}
C_{\epsilon}:=\sup_{R: \, p_{\rm out}(R,\snr)\leq \epsilon}R
\end{equation}
\end{definition}
\vspace*{0.2cm}

\begin{figure}[t]
\centering
\psfrag{1 block}{one block}
\psfrag{fb0}{${\rm FB}=0$}
\psfrag{fb1}{${\rm FB}=1$}
\psfrag{S2R}{${\rm S} \mbox{ with } 2R$}
\psfrag{R2R}{${\rm R} \mbox{ with } 2R$}
\includegraphics[width=0.33\textwidth]{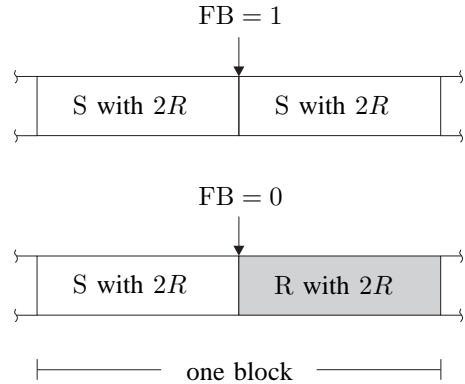}
\caption{Transmission model for incremental relaying. If the source-destination link is not in outage (feedback ${\rm FB}=1$), the source transmits during the second sub-block, too. If the source-destination link is in outage (feedback ${\rm FB}=0$), the relay aids communication during the second sub-block.}
\label{fig:blocks}
\end{figure}

\subsection{Incremental Relaying}

We consider incremental relaying (IR) as the cooperation protocol which exploits the availability of a one-bit feedback from the destination in the form of an ACK/NACK signal. Confer to Fig.~\ref{fig:blocks} for the following description. For a given $R$, we divide one transmission block into two sub-blocks of equal length. The transmission rate is set to $2R$ within each sub-block in order to have the same amount of information transmitted compared to the case where $\rm S$ transmits over the whole block with rate $R$. Now, if the source-destination link is not in outage at the end of first sub-block, information is transmitted over half a block and we get a rate of $2R$. Since we consider a block-fading model, this automatically means that during the second sub-block $\rm S$ can transmit its next message which will then be sent successfully to $\rm D$ and there is not need for the relay to aid communication. However, if the source-destination link is in outage, the relay transmits over the second sub-block and we get an overall rate of $R$. We define ${\cal A}$ as the event that the source-destination link is in outage, i.e., we have
\[{\cal A} := \{h_{\rm sd}: |h_{\rm sd}|^2 < \gamma \}, \]
where we use the definition
\begin{equation}
\gamma(R,\snr) := \frac{2^{2R} - 1}{\snr}
\end{equation}
and drop the dependence on $R$ and $\snr$ for the sake of brevity. In a similar fashion, we define
\begin{eqnarray}
{\cal B} &:=& \left\{h_{\rm sr}: |h_{\rm sr}|^2 < \gamma\right\} \nonumber \\
{\cal C} &:=& \left\{(h_{\rm sd},h_{\rm rd}): |h_{\rm sd}|^2 + |h_{\rm rd}|^2 < \gamma\right\}. \nonumber
\end{eqnarray}

The system is in outage either when both the source-destination as well as the source-relay link are in outage, or when the relay is able to decode, but the accumulation of SNR from the source and the relay at the destination still is not large enough to exceed a required minimum threshold. Dropping the dependence on $R$ and $\snr$ for simplicity, the outage probability can be written as:
\begin{eqnarray*}
p_{\rm out} \hspace*{-0.5em}&=&\hspace*{-0.5em} \Pr({\cal A})\Pr({\cal B})\Pr({\cal C}|{\cal AB}) + \Pr({\cal A})\Pr({\cal B}^c)\Pr({\cal C}|{\cal A}{\cal B}^c)  \\
\hspace*{-0.5em}&=&\hspace*{-0.5em} \Pr({\cal A})\Pr({\cal B}) + \Pr({\cal B}^c)\Pr({\cal C}),
\end{eqnarray*}
where ${\cal B}^c$ describes the complement of ${\cal B}$, $\Pr({\cal C}|{\cal AB})=1$, and $\Pr({\cal A})\Pr({\cal C}|{\cal A}{\cal B}^c)=\Pr({\cal C})$ due to ${\cal C} \subseteq {\cal A}$. With our system model, we get
\begin{eqnarray}
p_{\rm out} &=& \Pr\left(|h_{\rm sd}|^2<\gamma\right) \Pr\left(|h_{\rm sr}|^2<\gamma\right) \notag\\
&&+ \Pr\left(|h_{\rm sr}|^2\geq\gamma\right)\Pr\left(|h_{\rm sd}|^2+|h_{\rm rd}|^2<\gamma \right).
\end{eqnarray}
In order to be able to calculate the outage probability, we use the following lemma whose proof can be found in \cite{avestimehr07}.
\vspace*{0.2cm}
\begin{lemma}\label{lemma_cdf_general}
Let $w = \sum_{k=0}^{K}u_k$, where $u_k$ are independent exponentially distributed random variables with mean $\sigma_k^2$. If $g(x)$ is a continuous function at $x=0$ and $g(x) \rightarrow 0$ as $x \rightarrow 0$, then the cumulative distribution function $F$ of $w$ satisfies
\begin{equation}\label{eq:lemma_cdf_general}
  \lim\limits_{x \rightarrow 0} \frac{1}{g(x)^{K+1}} F(g(x)) = \frac{1}{(K+1)! \, \prod\limits_{k=0}^{K}\sigma_k^2}.
\end{equation}
\end{lemma}

\vspace*{0.2cm}
Outage probability in the low SNR regime can then be expressed as follows if the condition $\gamma \rightarrow 0$ for $\snr \rightarrow 0$ is met:
\begin{eqnarray}
\lim_{\substack{\epsilon \rightarrow 0 \\ \snr \rightarrow 0}}
\frac{p_{\rm out}}{\gamma^2} &=& \lim_{\substack{\epsilon \rightarrow 0 \\ \snr \rightarrow 0}}\left\{
\frac{\Pr(|h_{\rm sd}|^2<\gamma)}{\gamma}\frac{\Pr(|h_{\rm sr}|^2<\gamma)}{\gamma} \right. \nonumber \\
&& \left. + \frac{\Pr(|h_{\rm sd}|^2+|h_{\rm rd}|^2<\gamma)}{\gamma^2}\frac{\Pr(|h_{\rm sr}|^2\geq\gamma)}{1} \right\} \notag \\
&=& \frac{1}{\sigma_{\rm sd}^2}\frac{1}{\sigma_{\rm sr}^2} + \frac{1}{2\sigma_{\rm sd}^2\sigma_{\rm rd}^2} \cdot 1 \notag \\
&=& \frac{2\sigma_{\rm rd}^2+\sigma_{\rm sr}^2}{2\sigma_{\rm sd}^2\sigma_{\rm sr}^2\sigma_{\rm rd}^2}
\end{eqnarray}
Here, $\epsilon \rightarrow 0$ implies $\gamma \rightarrow 0$, which means that the rate is adapted in accordance to the SNR. From Definition \ref{def:out_cap}, the $\epsilon$-outage capacity is:
\begin{equation}\label{eq:c_e}
C_{\epsilon} = \frac{1}{2} \log_{2}\left( 1 + \snr \sqrt{\frac{2\sigma_{\rm sd}^2\sigma_{\rm sr}^2\sigma_{\rm rd}^2 \epsilon}{2\sigma_{\rm rd}^2+\sigma_{\rm sr}^2}} \right)
\end{equation}

\begin{figure}[t]
\centering
\psfrag{dsropt}{$d_{\rm sr}^{\ast}$}
\psfrag{0.32}{$0.32$}
\psfrag{0.36}{$0.36$}
\psfrag{0.4}{$0.4$}
\psfrag{0.44}{$0.44$}
\psfrag{alpha}{$\alpha$}
\psfrag{2}{$2$}
\psfrag{2.5}{$2.5$}
\psfrag{3}{$3$}
\psfrag{3.5}{$3.5$}
\psfrag{4}{$4$}
\psfrag{4.5}{$4.5$}
\psfrag{5}{$5$}
\includegraphics[width=0.5\textwidth]{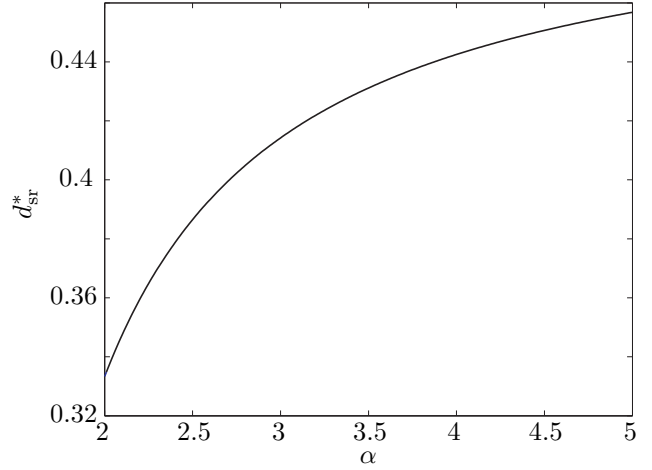}
\caption{Optimal source-relay distance $d_{\rm sr}^{\ast}$ versus path-loss factor $\alpha$. For $\alpha \rightarrow \infty$ $d_{\rm sr}^{\ast}$ tends asymptotically to $d_{\rm sr}^{\ast}=0.5$. $d_{\rm sr}^{\ast}(\alpha = 2)=1/3$.}
\label{fig:d_sr_opt_vs_path_loss_alpha}
\end{figure}

However, this expression does not include the variable transmission rate that occurs for incremental relaying in a long-term perspective. To account for that, we have to consider the average amount of sub-blocks required for transmission. If the source-destination link is not in outage, we need one sub-block, no matter if the relay is able to decode the source signal or not. If the source-destination link is in outage, we then have to transmit over two sub-blocks. Again, the number of sub-blocks required for transmission does not depend on the ability of the relay to decode the source signal. Let us define a random variable $N$ that denotes the number of transmission phases. The mean of $N$ becomes ${\mathbb E}(N) = 1 + \Pr(\cal A)$.
The \outcap of incremental relaying, denoted by the superscript ${\rm IR}$, can now be expressed as:
\begin{equation}\label{eq:c_e_ir}
C_{\epsilon}^{\rm IR} = \frac{2 C_{\epsilon}}{{\mathbb E}(N)}=\frac{1}{{\mathbb E}(N)}\log_{2}\left( 1 + \snr \sqrt{\frac{2\sigma_{\rm sd}^2\sigma_{\rm sr}^2\sigma_{\rm rd}^2 \epsilon}{2\sigma_{\rm rd}^2+\sigma_{\rm sr}^2}} \right)
\end{equation}
The factor $2/\mathbb{E}(N)$ is due to the fact of possible reduction of required transmission phases. If we only need one transmission phase, i.e., half a block (see Fig.~\ref{fig:blocks}), we have a gain of $2$. If we need two phases, i.e., the whole block, then we are as good as a relay network without feedback where the relay always transmits if it has been able to decode the source message. Therefore, we have
\begin{equation}
1 \leq \frac{C_{\epsilon}^{\rm IR}}{C_{\epsilon}} \leq 2.
\end{equation}
Assume that the relay is located on a straight line between the source and the destination. Accordingly, $d_{\rm rd}=1-d_{\rm sr}$. Moreover, let all distances be normalized to $d_{\rm sd}$ so that $\sigma_{\rm sd}^2=1$. This yields
\begin{equation}
C_{\epsilon}^{\rm IR}=\frac{1}{\mathbb{E}(N)} \log_{2}\left( 1 + \snr \sqrt{\frac{2\epsilon}{2d_{\rm sr}^{\alpha}+(1-d_{\rm sr})^{\alpha}}} \right).
\end{equation}
The optimal relay location which maximizes $\epsilon$-outage capacity becomes
\begin{equation}
d_{\rm sr}^{\ast}=\arg \max_{d_{\rm sr}} C_{\epsilon}^{\rm IR}=\arg \min_{d_{\rm sr}}\Psi(d_{\rm sr}),
\end{equation}
where $\Psi(d_{\rm sr})=2d_{\rm sr}^{\alpha}+(1-d_{\rm sr})^{\alpha}$. It can easily be seen that the optimal relay location is independent of $\snr$ and $\epsilon$. By setting the derivation of $\Psi(d_{\rm sr})$ with respect to $d_{\rm sr}$ equal to $0$,
\begin{equation}
\frac{\partial \Psi(d_{\rm sr})}{\partial d_{\rm sr}} = 0,
\end{equation}
we get
\begin{equation}
d_{\rm sr}^{\ast}=\frac{1}{1 + \sqrt[\alpha-1]{2}} < 0.5. 
\end{equation}
The fact that $d_{\rm sr}^{\ast}$ is bounded by $0.5$ corresponds to results presented in \cite{kramer05}, where it is demonstrated that decode-and-forward performs better if the relay is located closer to the source. For free-space propagation, e.g., we have
$d_{\rm sr}^{\ast}(\alpha=2)=1/3$ and for $\alpha=3$, $d_{\rm sr}^{\ast}(\alpha=3)=\sqrt{2} - 1 \approx 0.4142$. This is illustrated in Fig.~\ref{fig:d_sr_opt_vs_path_loss_alpha}. We clearly see that $d_{\rm sr}^{\ast}$ is monotonically increasing in $\alpha$. For the worst channel condition, i.e., $\alpha \rightarrow \infty$, the relay should be located half-way between source and destination, which also seems clear from an intuitive point of view.

\subsection{Cut-Set Bound}
We next consider the cut-set bound (max-flow min-cut theorem) of the relay channel with Gaussian codebooks. Since it is an upper bound on the flow of information in any network that consists of multiple terminals, it clearly serves as an upper bound for incremental relaying. Hence, the best we could do is to achieve the cut-set bound. The cut-set bound of the relay channel yields:
\begin{eqnarray}
I &=& \min\{ \underbrace{\log_{2}(1+(|h_{\rm sd}|^2+|h_{\rm sr}|^2)\snr)}_{\text{BC-cut}},\nonumber \\
&&\underbrace{\log_{2}(1+(|h_{\rm sd}|^2+|h_{\rm rd}|^2)\snr)}_{\text{MAC-cut}} \}
\end{eqnarray}
The BC-cut and the MAC-cut are illustrated in Fig.~\ref{fig:network}. We now follow exactly the same steps that we used for incremental relaying in order to get an expression of $\epsilon$-outage capacity. First, outage probability in the low SNR regime, where again the condition $\gamma \rightarrow 0$ for $\snr \rightarrow 0$ must be met, becomes:
\begin{eqnarray}
\lim_{\substack{\epsilon \rightarrow 0 \\ \snr \rightarrow 0}}
\frac{p_{\rm out}}{\gamma^2} &=& \lim_{\substack{\epsilon \rightarrow 0 \\ \snr \rightarrow 0}}\left\{
\frac{\Pr(|h_{\rm sd}|^2+|h_{\rm sr}|^2<\gamma)}{\gamma^2} \right. \nonumber \\
\hspace*{-1.5cm} && \hspace*{-1.5cm} \left. + \frac{\Pr(|h_{\rm sd}|^2+|h_{\rm sr}|^2\geq\gamma)}{1}\frac{\Pr(|h_{\rm sd}|^2+|h_{\rm rd}|^2<\gamma)}{\gamma^2} \right\} \notag \\
&=& \frac{\sigma_{\rm rd}^2+\sigma_{\rm sr}^2}{2\sigma_{\rm sd}^2\sigma_{\rm sr}^2\sigma_{\rm rd}^2}
\end{eqnarray}
Then, the \outcap is
\begin{equation}\label{eq:c_e_csb}
C_{\epsilon}^{\rm CSB} \geq \frac{1}{1+\epsilon}\log_{2}\left( 1 + \snr \sqrt{\frac{2\sigma_{\rm sd}^2\sigma_{\rm sr}^2\sigma_{\rm rd}^2 \epsilon}{\sigma_{\rm rd}^2+\sigma_{\rm sr}^2}} \right),
\end{equation}
where the superscript $\rm CSB$ stands for cut-set bound and we have applied $\mathbb{E}(N) \leq 1 + \epsilon$. This upper bound is reasonable since our aim is to have an overall outage probability lower then or equal to $\epsilon$. The probability for an outage after the first sub-block clearly is higher than $\epsilon$, i.e., $\Pr({\cal A}) \geq \epsilon$, and we get a tighter upper bound by setting $\mathbb{E}(N) \leq 1 + \epsilon$.


\section{Comparison}\label{sec:comp}
\subsection{One-relay Case}
In this section, we compare the \outcap of incremental relaying to the cut-set bound. For that purpose, we use the following performance criterion.

\vspace*{0.2cm}
\begin{definition}
The ratio between incremental relaying and the cut-set bound for the same value of $\epsilon$ is defined as
\begin{equation}
\Delta(\epsilon)=\frac{C_{\epsilon}^{\rm IR}}{C_{\epsilon}^{\rm CSB}}.
\end{equation}
Since the cut-set bound is an upper bound that describes the maximal achievable rate in a network, it is obvious that $\Delta(\epsilon) \leq 1$.
\end{definition}
\vspace*{0.2cm}

Applying \eqref{eq:c_e_ir} and \eqref{eq:c_e_csb}, we get
\begin{equation}
\Delta(\epsilon)\leq \sqrt{\frac{\sigma_{\rm rd}^2+\sigma_{\rm sr}^2}{2\sigma_{\rm rd}^2+\sigma_{\rm sr}^2}} = \sqrt{\frac{\left(\frac{d_{\rm sr}}{d_{\rm rd}}\right)^{\alpha}+1}{2\left(\frac{d_{\rm sr}}{d_{\rm rd}}\right)^{\alpha}+1}},
\end{equation}
where we used $\ln(1+x) \approx x$ for small values of $x$ and $\Pr({\cal A}) \approx \epsilon$. We now see that $\Delta(\epsilon) \in [1/\sqrt{2},1]$ for $\gamma \rightarrow 0$. The lower bound ($1/\sqrt{2}$) describes the case when the relay is placed close to the destination, whereas the upper bound ($1$) represents the case when the relay is located close to the source.

\begin{figure}[t]
\centering
\psfrag{0}{$0$}
\psfrag{0.1}{$0.1$}
\psfrag{0.2}{$0.2$}
\psfrag{0.3}{$0.3$}
\psfrag{0.4}{$0.4$}
\psfrag{0.5}{$0.5$}
\psfrag{0.6}{$0.6$}
\psfrag{0.7}{$0.7$}
\psfrag{0.75}{$0.75$}
\psfrag{0.8}{$0.8$}
\psfrag{0.85}{$0.85$}
\psfrag{0.9}{$0.9$}
\psfrag{0.95}{$0.95$}
\psfrag{1}{$1$}
\psfrag{0.35}{$0.35$}
\psfrag{0.45}{$0.45$}
\psfrag{DeltaCe}{$\Delta(\epsilon)$}
\psfrag{upper bound}{upper bound}
\psfrag{lower bound}{lower bound}
\psfrag{e=10(-3)}{$\epsilon = 10^{-3}$}
\psfrag{e=10(-4)}{$\epsilon = 10^{-4}$}
\psfrag{e=10(-5)}{$\epsilon = 10^{-5}$}
\psfrag{dsr}{$d_{\rm sr}$}
\includegraphics[width=0.5\textwidth]{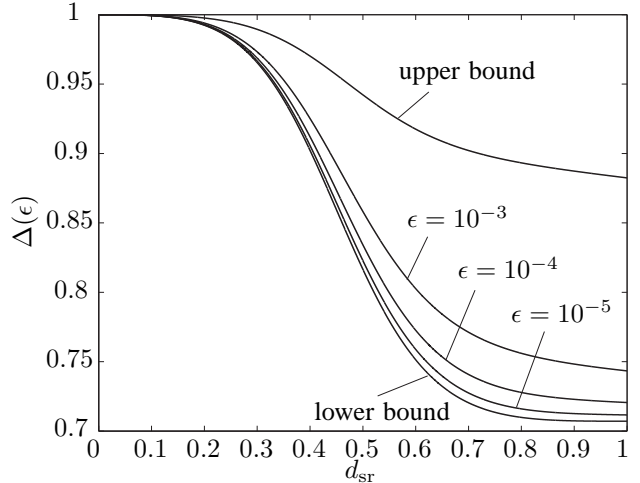}
\caption{Ratio $\Delta(\epsilon)$ of \outcap of incremental relaying to the cut-set bound for $\gamma \rightarrow 0$ and $\alpha = 3$. When the relay is placed close to the destination, we get $\Delta(\epsilon)=1/\sqrt{2}$. When the relay is located close to the source, we have $\Delta(\epsilon)=1$.}
\label{fig:outage_capacity_ir_vs_df}
\end{figure}

Fig.~\ref{fig:outage_capacity_ir_vs_df} illustrates the ratio $\Delta(\epsilon)$ of \outcap of incremental relaying to the cut-set bound for $\gamma \rightarrow 0$. When the relay is placed close to the destination, we get the poorest performance of incremental relaying. When the relay is located close to the source, incremental relaying with decode-and-forward is optimal.

\subsection{Extension to $K$ Relays}

The calculation of the outage probability of incremental relaying for an arbitrary number of relays is involved. Normally, one would have to investigate all possibilities how information could be sent from the source over the relays to the destination. In a network with $K$ relays, this leads to $2^K$ different cuts. We simplify the calculation by assuming that either all $K$ relays can decode the source message or not. The outage probability can then be lower bounded by
\begin{eqnarray}
p_{\rm out} &\geq& \Pr(|h_{\rm sd}|^2<\tilde{\gamma}) \prod_{k=1}^{K}\Pr(|h_{{\rm sr}_k}|^2<\tilde{\gamma}) \nonumber \\
&&+ \prod_{k=1}^{K}\Pr(|h_{{\rm sr}_k}|^2\geq\tilde{\gamma})\Pr(|h_{\rm sd}|^2 + \sum_{k=1}^{K}|h_{{\rm r}_k{\rm d}}|^2<\tilde{\gamma}), \nonumber
\end{eqnarray}
where
\begin{equation}
\tilde{\gamma} = \frac{2^{(K+1)R} - 1}{\snr}.
\end{equation}
We see that now the source has to transmit with an initial rate of $(K+1)R$. By applying Lemma \ref{lemma_cdf_general}, we get:
\begin{equation}
\lim_{\substack{\epsilon \rightarrow 0 \\ \snr \rightarrow 0}}
\frac{p_{\rm out}}{\tilde{\gamma}^{K+1}} \geq \frac{(K+1)! \prod_{k=1}^{K}\sigma_{{\rm r}_k{\rm d}}^2 + \prod_{k=1}^{K}\sigma_{{\rm sr}_k}^2}{(K+1)! \sigma_{\rm sd}^2 \prod_{k=1}^{K}\sigma_{{\rm r}_k{\rm d}}^2\sigma_{{\rm sr}_k}^2}
\end{equation}
This leads to an upper bound on the \outcap of
\begin{eqnarray}\label{eq:c_e_ir_k}
C_{\epsilon}^{\rm IR} &\leq&   \frac{1}{{\mathbb E}_K(N)}   \nonumber \\
\hspace*{-1.4cm}&&\hspace*{-1.4cm} \times \log_2\left( 1 + \snr \sqrt[K+1]{\frac{(K+1)! \sigma_{\rm sd}^2 \prod_{k=1}^{K}\sigma_{{\rm r}_k{\rm d}}^2\sigma_{{\rm sr}_k}^2 \epsilon}{(K+1)! \prod_{k=1}^{K}\sigma_{{\rm r}_k{\rm d}}^2 + \prod_{k=1}^{K}\sigma_{{\rm sr}_k}^2}} \right), \nonumber
\end{eqnarray}
where ${\mathbb E}_K(N) = 1 + \sum_{k=1}^{K} \Pr({\cal C}_k)$ and
\begin{equation}
{\cal C}_k = \{ |h_{\rm sd}|^2 + \sum_{l=1}^{k-1}|h_{{\rm r}_l{\rm d}}|^2 < \tilde{\gamma} \}.
\end{equation}
The event ${\cal C}_k$ describes the accumulation of SNR at the destination which is caused by the fact that the relays transmit in a successive manner. This means that if the accumulated SNR of the source's and the first relay's transmission is not enough for the destination to decode, then the second relay transmits. If the accumulated SNR of the source's, the first, and the second relay's transmission is not enough, the third relay transmits, and so on.

For the cut-set bound, an upper bound on mutual information is given by
\begin{eqnarray}
I &\leq& \min\{ \underbrace{\log_{2}(1+(|h_{\rm sd}|^2+\sum_{k=1}^{K}|h_{{\rm sr}_k}|^2)\snr)}_{\text{BC-cut}},\nonumber \\
&&\underbrace{\log_{2}(1+(|h_{\rm sd}|^2+\sum_{k=1}^{K}|h_{{\rm r}_k{\rm d}}|^2)\snr)}_{\text{MAC-cut}} \},
\end{eqnarray}
where we only considered the BC-cut and the MAC-cut and neglected any mix-terms. With
\begin{equation}
\lim_{\substack{\epsilon \rightarrow 0 \\ \snr \rightarrow 0}}
\frac{p_{\rm out}}{\tilde{\gamma}^{K+1}} \leq \frac{\prod_{k=1}^{K}\sigma_{{\rm r}_k{\rm d}}^2 + \prod_{k=1}^{K}\sigma_{{\rm sr}_k}^2}{(K+1)! \sigma_{\rm sd}^2 \prod_{k=1}^{K}\sigma_{{\rm r}_k{\rm d}}^2\sigma_{{\rm sr}_k}^2}
\end{equation}
\outcap then becomes
\begin{eqnarray}\label{eq:c_e_csb_k}
C_{\epsilon}^{\rm CSB} &\geq& \frac{1}{1+K\epsilon} \nonumber \\
\hspace*{-1.2cm}&&\hspace*{-1.2cm}\times \log_2\left( 1 + \snr \sqrt[K+1]{\frac{(K+1)! \sigma_{\rm sd}^2 \prod_{k=1}^{K}\sigma_{{\rm r}_k{\rm d}}^2\sigma_{{\rm sr}_k}^2 \epsilon}{\prod_{k=1}^{K}\sigma_{{\rm r}_k{\rm d}}^2 + \prod_{k=1}^{K}\sigma_{{\rm sr}_k}^2}} \right). \nonumber
\end{eqnarray}


\vspace*{0.5em}
\section{Conclusions and Further Research}\label{sec:con}
In this paper we derived the \outcap of incremental relaying when repetition coding, which is optimal in the low SNR regime, is applied and the relays perform decode-and-forward. We showed that the optimal relay location which maximizes \outcap only depends on the channel parameters and is independent of SNR and outage probability. We compared our results to the cut-set bound and demonstrated that the ratio between the \outcap of incremental relaying and the cut-set bound lies within $1/\sqrt{2}$ and $1$ in the best case scenario. Lastly, we derived lower bounds on the \outcap for the case of $K$ relays for incremental relaying and the cut-set bound. In \cite{avestimehr07} it has been shown that a bursty version of amplify-and-forward (BAF) is rate-optimal for low values of SNR for a frequency division communication model. The performance of the BAF strategy for incremental relaying is currently under investigation. This also raises questions with respect to relay selection and power allocation. Those issues are particularly important for energy-constrained networks such as ad-hoc and sensor networks.

%

%
%

\end{document}